\documentstyle[amssymb,amsmath,epsf,twocolumn,fleqn]{article}

\begin{document}
\renewcommand{\textfraction}{0.1}
\renewcommand{\topfraction}{0.8}
\rule[-8mm]{0mm}{8mm}
\begin{minipage}[t]{15cm}
\begin{center}
{\LARGE \bf Peierls-insulator Mott-insulator transition in 1D}\\[4mm] 
H.~Fehske$^{\rm a}$, G.~Wellein$^{\rm b}$, A. Wei{\ss}e$^{\rm a}$,
F. G\"ohmann$^{\rm a}$, H. B\"uttner$^{\rm a}$, and 
A. R. Bishop$^{\rm c}$\\[3mm]
$^{\rm a}${Physikalisches Institut, Universt\"at Bayreuth, 
95440 Bayreuth, Germany}\\
$^{\rm b}${Regionales Rechenzentrum Erlangen, Universit\"a{}t Erlangen, 
  91058 Erlangen, Germany}\\
$^{\rm c}$
{MSB262,  Los Alamos National Laboratory, Los Alamos, New Mexico 87545, 
U.S.A.}\\[4.5mm]
\end{center}
{\bf Abstract}\\[0.2cm]
\hspace*{0.5cm}
In an attempt to clarify the nature of the crossover 
from a Peierls band insulator to a Mott 
Hubbard insulator, we analyze ground-state and spectral 
properties of the one-dimensional half-filled Holstein 
Hubbard model using exact diagonalization techniques.
\\[0.2cm]
\end{minipage}\\[4.5mm]
\normalsize
In a wide range of quasi-one-dimensional materials, such 
as MX chains, conjugated polymers or ferroelectric perovskites, 
the itineracy of the electrons strongly competes with
electron-electron and electron-phonon (EP) interactions, which
tend to localize the charge carriers by establishing
spin-density-wave and charge-density-wave ground states,
respectively. Hence, at half-filling, Peierls (PI) or Mott (MI) 
insulating phases are energetically favored 
over the metallic state. An interesting and still 
controversial question is whether or not  only 
one quantum critical point separates the PI and MI 
phases at $T=0$~\cite{FGN99}.
Furthermore, how is the crossover modified 
when phonon dynamical effects, which are known to be 
of particular importance in low-dimensional materials~\cite{JZW99,FHW00}, 
are taken into account?

The paradigm in studies of this subject is the half-filled
Holstein-Hubbard model (HHM), defined by the Hamiltonian 
\begin{eqnarray}
{\cal H}\!\!\!\!&=&\!\!\!\!
-t\sum_{i,\sigma}(c^{\dagger}_{i\sigma}c^{}_{i+1 \sigma}
   \!+\!\mbox{H.c.})+U\sum_i n_{i\uparrow}n_{i\downarrow}
\nonumber\\
&& + g\omega_0\sum_{i,\sigma} (b_i^{\dagger} + b_i^{}) 
n_{i\sigma}+ \omega_0\sum_{i} b_i^{\dagger} b_i^{}.
\label{phm}
\end{eqnarray}
Here $c^{\dagger}_{i\sigma}$ creates a spin-$\sigma$ 
electron at Wannier site~$i$
($n_{i,\sigma}=c^{\dagger}_{i\sigma}c^{}_{i\sigma}$), 
$b_i^{\dagger}$ creates a local phonon
of frequency $\omega_0$, $t$ denotes the hopping integral,
$U$ is the on-site Hubbard repulsion,  
$g$ is a measure of  the 
\begin{figure}[t]
\unitlength1mm
\begin{picture}(70,64)
\end{picture}
\end{figure}
EP coupling strength, and the summation over $i$ extends over a periodic
chain of $N$ sites.

Applying basically exact numerical methods~\cite{WFWB00}, 
we are able to diagonalize the HHM on finite chains, preserving 
the full dynamics of the phonons. 
In order to characterize the ground-state and spectral properties 
of the HHM in different parameter regimes, we have calculated 
the charge- and spin-structure factors at $q=\pi$   
\begin{eqnarray}
S_c(\pi)\!\!&=&\!\!\frac{1}{N}\sum_{j,\sigma\sigma'} (-1)^j
(\langle n_{i\sigma}n_{i+j\,\sigma'}\rangle -\tfrac{1}{4})\\
S_s(\pi)\!\!&=&\!\!\frac{1}{N}\sum_j (-1)^j\langle S_i^zS_{i+j}^z\rangle
\label{scs}
\end{eqnarray}
($S_i^z=(n_{i\uparrow}-n_{i\downarrow})/2$), the local magnetic
moment $L=3\langle (S_i^z)^2\rangle$, the kinetic energy 
$E_{kin}=-t\langle \sum_{i,\sigma}(c^{\dagger}_{i\sigma}c^{}_{i+1 \sigma}
   +\mbox{H.c.})\rangle $, and the incoherent part of the optical
conductivity      
\begin{equation}  \label{sigmareg}              
\hspace*{-0.7cm}\sigma^{\rm reg}(\omega)\!=\!\frac{\pi}{N}\!\sum_{m\neq 0}\!
\frac{|\langle\psi_0|\hat{j}
|\psi_m\rangle|^2}{E_m-E_0}\,\delta[\omega\!-\!E_m\!+\!E_0], 
\end{equation} 
where $\hat{j}=-{\rm i}e t\sum_{i\sigma}(c_{i\sigma}^{\dagger}
c_{i+1\,\sigma}^{}-c_{i+1\,\sigma}^{\dagger}c_{i\sigma}^{})$. Some 
typical results are shown in Figs.~1 and~2.\\[0.2cm]
Our conclusions can be summarized as follows:

(i) At $U=0$ the ground-state is a Peierls distorted state
in the adiabatic limit $\omega_0\to 0$ for any finite EP coupling.  
As in the Holstein model of spinless fermions~\cite{FHW00}, 
at $\omega_0>0$ quantum phonon fluctuations destroy the Peierls instability 
for small EP interaction strength $g$~\cite{JZW99}.   
Above a critical threshold $g_c(\omega_0)$, the HHM
describes a PI with gapped spin and charge excitations. 
In the non-adiabatic strong EP coupling regime, 
the system is typified by a charge-ordered bipolaronic
insulator rather than a traditional Peierls band insulator.
\begin{figure}[t]
\centerline{\mbox{\epsfxsize 7.5cm\epsffile{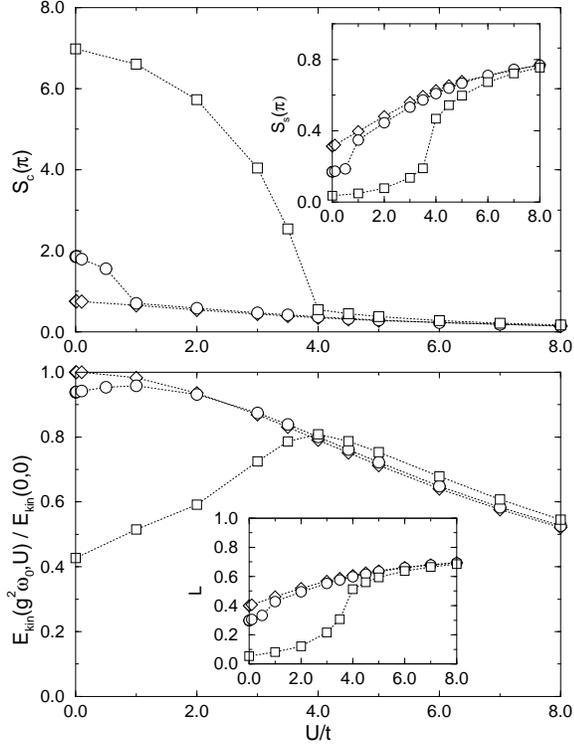}}}
\caption{Staggered charge- and spin-density correlations (upper panel),
kinetic energy and local magnetic moment (lower panel) 
in the ground state of the Holstein-Hubbard model ($\omega_0/t=1$; $N=8$). 
Results are shown at $g^2\omega_0/t=0$ (diamonds), 0.5 (circles), and
2.0 (squares).}
\end{figure} 
The PI regime is characterized by a large (small) charge (spin) 
structure factor, a strongly reduced kinetic energy, and an 
optical response that is dominated by multiphonon absorption 
and emission processes.

(ii) Increasing $U$ at fixed $g$,  the Peierls dimerization and the 
concomitant charge order are suppressed. 
Accordingly the system evolves from the PI 
to the MI regime. From our numerical data we found evidence 
for {\it only one} critical point $U_c$ 
(cf., e.g., the development of the optical gap in the 
conductivity spectra shown Fig.~2). At $U_c$, in our 
finite system a site parity change of the ground
\begin{figure}[t]
\centerline{\mbox{\epsfxsize 8cm\epsffile{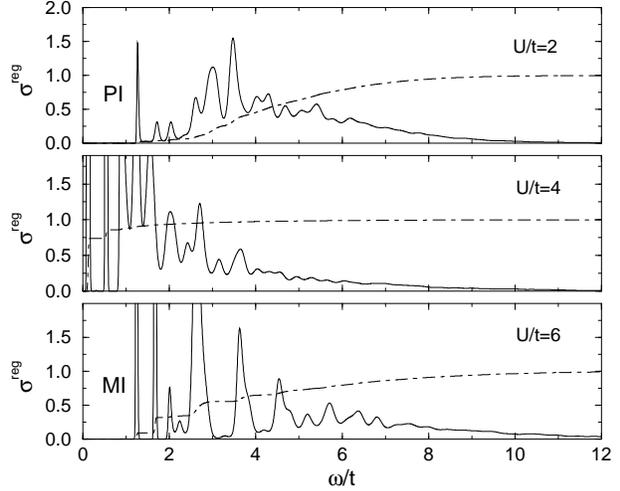}}}
\caption{\small Optical absorption in the HHM (\mbox{$\omega_0/t\!=\!1$}; 
$N=8$).
Dashed lines give the integrated spectral weights 
$S^{reg}(\omega)=\int_0^\omega \sigma^{reg}(\omega')\,d\omega'$
[normalized by $S^{reg}(\infty)$].}
\end{figure}
\begin{table}[h!]
\centerline{\begin{tabular}{c|ccc}
\rule{0mm}{4mm}$U/t$&2&4&6\\[0.1cm] \hline
\rule{0mm}{4mm}$P$ & $+$ & $-$ & $-$\\[0.1cm] 
\rule{0mm}{4mm}$\Delta_c(8)/t$ & 1.84 & 0.24 & 1.7\\[0.1cm] 
\rule{0mm}{4mm}$\Delta_s(8)/t$ & 1.62 & 0.13 & 0.28
\end{tabular}}
\caption{Parity of the ground state of the HHM, 
\protect{$P|\psi_0\rangle=\pm |\psi_0\rangle$}, where 
the site inversion symmetry operator $P$ is defined by  
\protect{$Pc_{i\sigma}^\dagger P^\dagger=c_{N-i\sigma}^\dagger$} for 
\protect{$i=0,1,\ldots,N-1$}. 
Charge and spin gaps, defined by \protect{
$\Delta_c(N)=E_0(\tfrac{N}{2}+1,\tfrac{N}{2})+E_0(\tfrac{N}{2}
-1,\tfrac{N}{2})-2E_0(\tfrac{N}{2},\tfrac{N}{2})$} and
\protect{$\Delta_s(N)=E_0(\tfrac{N}{2}+1,\tfrac{N}{2}-1)
-E_0(\tfrac{N}{2},\tfrac{N}{2})$}, respectively, where 
$E_0(N_{\uparrow},N_{\downarrow})$ denotes the ground-state energy 
of the system with $N_{\uparrow}$ spin-up and $N_{\downarrow}$ 
spin-down electrons. Note that $\Delta_c$ incorporates 
ground-state lattice relaxation effects and therefore differs from 
the optical gap~\protect\cite{JZW99}.
In the infinite system, we expect that 
$\Delta_c=\Delta_s\geq 0$ for $U\leq U_c$, whereas  
$\Delta_c>\Delta_s=0$ for $U> U_c$. 
Results are given at $g^2\omega_0=2,\;\omega_0/t=1$.}
\end{table}
state takes place
from $P=+1$ (PI) to  $P=-1$ (MI), and both 
the spin and the charge gaps are expected to vanish in 
the thermodynamic limit (cf. Tab.~I). 

(iii) Above $U_c$, in the MI phase, the low-energy physics of the system
is governed by gapless spin and massive charge excitations.
In the Mott Hubbard insulating regime the optical gap is by its nature
a correlation gap. It is rapidly destroyed by doping the system 
away from half filling~\cite{Je98}.
\end{document}